\documentclass[%
 superscriptaddress,
 reprint,
 showpacs,preprintnumbers,
 nofootinbib,
 amsmath,amssymb,
 aps,
 prc,
 floatfix,
]{revtex4-1}
\usepackage{graphicx}% Include figure files

\usepackage{amssymb,amsmath,amstext,amsthm,amsfonts}
\usepackage[utf8]{inputenc}
\usepackage{float}
\usepackage{url}
\usepackage{soul}

\newcommand{\pt}{p_\textrm{T}}
\newcommand{\plong}{p_\textrm{L}}

\newcommand{\dpt}{\delta \pt}
\newcommand{\dplong}{\delta \plong}
\newcommand{\dptt}{\delta p_\textrm{TT}}

\usepackage{lineno}
\begin{document}

	\title{A new method for an improved anti-neutrino energy reconstruction with charged-current interactions in next-generation detectors}

	\author{L. Munteanu}
	\affiliation{IRFU, CEA Saclay, Gif-sur-Yvette 91190, France}
	
	\author{S. Suvorov}
	\affiliation{IRFU, CEA Saclay, Gif-sur-Yvette 91190, France}
	\affiliation{Institute for Nuclear Research of the Russian Academy of Sciences, Moscow, Russia}	

	\author{S. Dolan}
	\email[Contact e-mail: ]{Stephen.Joseph.Dolan@cern.ch}
	\affiliation{CERN, Switzerland}

  	\author{D. Sgalaberna}
 			\email[Contact e-mail: ]{Davide.Sgalaberna@cern.ch}
	\affiliation{CERN, Switzerland}
 
	\author{S. Bolognesi}
	\affiliation{IRFU, CEA Saclay, Gif-sur-Yvette 91190, France}
	
	\author{S. Manly}
	\affiliation{University of Rochester, USA}
	
	\author{G. Yang}
	\affiliation{Stony Brook University, USA}

	\author{C. Giganti}
	\affiliation{LPNHE, Unversite` Sorbonne, France}

	\author{K. Iwamoto}
	\affiliation{University of Tokyo, Japan}
	
	\author{C. Jesús-Valls}
	\affiliation{Institut de Fisica d’Altes Energies (IFAE), The Barcelona Institute of Science and Technology, Campus UAB, Bellaterra (Barcelona), Spain}

\begin{abstract}
\noindent 
We propose and validate a method of anti-neutrino energy reconstruction for charged-current meson-less interactions on composite fully active targets containing hydrogen (such as hydrocarbon scintillator), which is largely free of the poorly understood nuclear effects that usually distort and bias attempts to measure neutrino energy. The method is based on the precise event-by-event measurement of the outgoing neutron kinetic energy and the subsequent assessment of the momentum imbalance on the plane transverse to the incoming anti-neutrino direction. For an anti-neutrino flux peaked at around 600 MeV measured using a finely grained  $2\times2\times2$ m$^3$ 3D scintillator tracker the neutrino energy resolution is expected to be around 7\%, compared to the 15\% expected using traditional neutrino energy reconstruction techniques. Analogous results can be obtained for other detectors with similar characteristics.

\end{abstract}

\maketitle
\section{Introduction}
\label{sec:introduction}

Current and future long baseline neutrino oscillation experiments~\cite{Abe:2011ks,Abe:2016tez,Abe:2019whr,Ayres:2007tu,Acciarri:2015uup,Abe:2018uyc} use GeV-scale (anti)neutrino beams to study neutrino oscillations by analysing how the beam's oscillated flavour content changes as a function of neutrino energy at a detector placed at some distance from the neutrino beam production point. In order to extract the oscillation parameters, the experiments measure the neutrino-nucleus interaction rate of a particular neutrino flavour as a function of some observable(s) which can be related to the true incoming neutrino energy. A particular difficulty for neutrino oscillation experiments is the accurate inference of the true neutrino energy from what is measured at the detectors~\cite{Alvarez-Ruso:2017oui,Katori:2016yel}. This is especially important for future experiments, where the evaluation of the CP-violating phase ($\delta_{CP}$) will require a precision measurement of the shape of the neutrino energy spectrum.

The T2K~\cite{Abe:2011ks} and Hyper-K~\cite{Abe:2018uyc} experiments reconstruct the neutrino energy from the measured lepton kinematics following a meson-less neutrino interaction, assuming the interactions are charged-current quasi-elastic (CCQE) scatters off a single stationary nucleon with some fixed binding energy to its parent nucleus (kinematic energy reconstruction)~\cite{Abe:2017uxa}. Experiments with far-detectors capable of calorimetry, such as at NOvA~\cite{NOvA:2018gge} and DUNE~\cite{Acciarri:2015uup} reconstruct the neutrino energy by summing all energy deposited within the detector (calorimetric energy reconstructions) whilst also measuring the momentum of the tracked particles by the distance they travel.

Both of these methods are subject to significant potential biases. The kinematic energy reconstruction does not account for the possibility of having observed a meson-less inelastic interactions such as those in which a neutrino scatters off a bound state of two nucleons (2p2h interactions) or those that produce a pion which is either not reconstructed or absorbed through hadronic re-scattering inside the nuclear medium (known as final state interactions, FSI). Similarly the calorimetric method does not account for the energy carried away by neutrons which are not detected. Moreover, both methods must contend with the smearing of neutrino energy reconstruction coming from the non zero initial state target nucleon's momentum (Fermi motion) and binding energy. These nuclear effects are poorly understood and models of them can vary dramatically in their predictions~\cite{Alvarez-Ruso:2017oui,Katori:2016yel}. This leads to an unknown bias in neutrino energy reconstruction and therefore also in the extracted neutrino oscillation parameters. Furthermore, the extrapolation of constraints from a near detector to a far detector in neutrino oscillation experiments is also strongly dependent on these nuclear effects, even if they have identical target nuclei and acceptances. Due to oscillations, the energy spectrum is different in the near and far sites and so near-detector constraints on the unoscillated flux and neutrino interactions cannot be applied without using a neutrino interaction model. The relation between the reconstructed and true neutrino energy is therefore different at the near and far detectors. 

If a sample of neutrino interactions weakly affected by nuclear effects could be identified, then the reconstructed neutrino energy maps more accurately to the true neutrino energy and the mapping becomes more similar at the near and far detectors. Such a sample of events could also be used to infer the neutrino flux at the near detector, in a way that has little dependence on the details of the neutrino interaction model. It has been shown that this can partially be achieved by analysing the momentum imbalance between the hadronic and leptonic parts of the neutrino interaction~\cite{Lu:2015hea,Furmanski:2016wqo,Duyang:2019prb}. These methods typically rely on identifying neutrino interactions with hydrogen, which are not subject to nuclear effects at all (if the neutrino energy was to be inferred from interactions off a hydrogen target the energy reconstruction would be perfect before considering the detector resolution). Moreover, if neutrino cross sections could be measured on both hydrogen and a heavier target, the difference would offer a direct probe of nuclear effects. However, the proposed methods largely rely on either having a hydrogen target detector or on subtracting a poorly understood nuclear background from a composite target which would be difficult to achieve. In this manuscript we demonstrate a method of identifying a sample of anti-neutrino interactions which are almost free of nuclear effects and that can be applied to fully-active plastic scintillator detectors which are currently in development~\cite{Sgalaberna:2017khy}. Although a perfect detector could extract a pure hydrogen sample, any realistic application of this technique will also contain interactions on a nuclear target but importantly these interactions will be those which are themselves minimally impacted by nuclear effects. Similarly to other methods, this is achieved by measuring the imbalance between the final-state lepton and neutron on the plane transverse to the incoming neutrino.

We begin by explaining our method of neutrino energy reconstruction and how we simulate its application to neutrino interactions which are subject to realistic detector smearing and acceptances. We then explore how this method could be used to improve neutrino energy reconstruction and lower key systematics in future neutrino oscillation analyses.

\section{Methodology}
\label{sec:methodology}

\subsection{Transverse momentum imbalance}
\label{sec:dpt}

Nuclear effects have previously been studied by analysing the kinematic imbalance of outgoing particles in the plane transverse to the incoming neutrino direction~\cite{Lu:2015hea, Lu:2015tcr, Abe:2018pwo, Lu:2018stk, Dolan:2018sbb, Dolan:2018zye, Lu:2019nmf, Dolan:2019bxf, neutNewPaperInProgress}. In past analyses this has only been measured using a neutrino beam, as the final state protons are relatively easy to measure. Here, we consider the case of the CCQE anti-neutrino interaction: $\bar{\nu} p \rightarrow \mu^+ n$ with a neutron in the final state. In these interactions the momentum imbalance can be simply defined as:
\begin{align}
\dpt &= |\overrightarrow{p}_T^l + \overrightarrow{p}_T^n|, 
\label{eq:dpt}
\end{align}

where $p^n$ and $p^l$ are the outgoing neutron and lepton momenta, and the $T$ index is the projection of the vector on the plane transverse to the incoming neutrino direction. For such interactions on a free nucleon (a hydrogen target) the transverse momentum of the final state is balanced and so $\dpt$ vanishes. On the other hand, interactions on nuclear targets are subject to Fermi motion, binding energy and FSI and thus $\dpt$ becomes non-zero. Furthermore, even with a perfect detector such CCQE interactions on a nuclear target are not experimentally accessible as additional particles in the final state can be absorbed inside the nuclear medium by FSI processes. The closest experimentally accessible topology are interactions in which there are no observed pions (or heavier mesons) in the final state (CC0$\pi$ interactions), but these contain a non-negligible portion of CCnonQE interactions such as 2p2h and pion-production (and then absorption by FSI).  

\begin{figure}
\centering
\includegraphics[width=10cm]{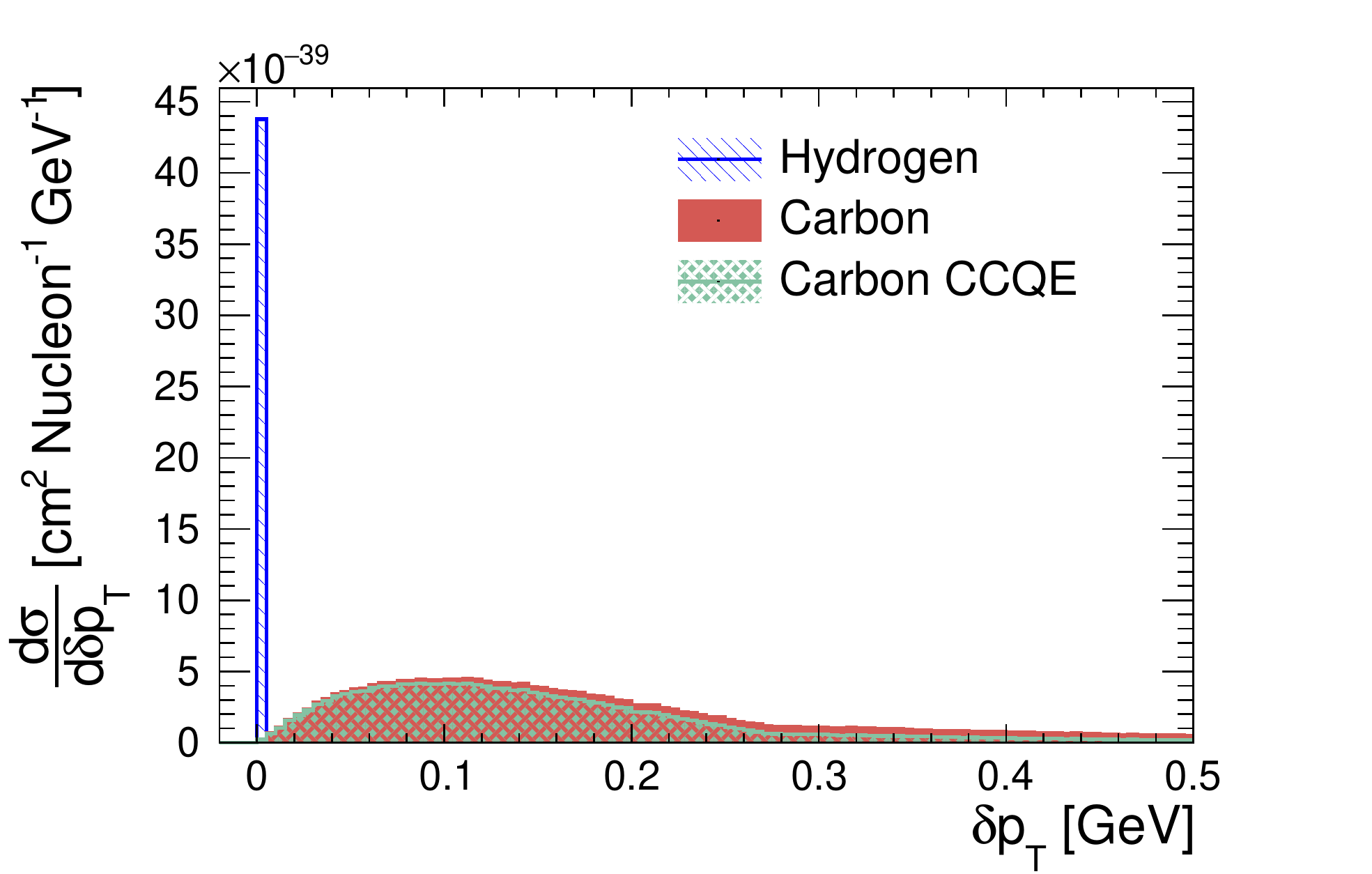}
\caption{\label{fig:dpt-no-smearing} The differential cross-section for CC0$\pi$ interactions on a hydrocarbon target as a function of $\delta p_T$ for anti-neutrino interactions from the T2K experiment's anti-neutrino flux~\cite{Abe:2012av,t2kfluxurl} according to the NEUT 5.4.0 simulation. The cross-section is split by the target nucleus and whether or not the interaction is CCQE or not (only for carbon, as all CC0$\pi$ interactions on hydrogen are CCQE). No detector smearing or acceptance effects are considered.}
\end{figure}

\begin{figure}
\centering
\includegraphics[width=10cm]{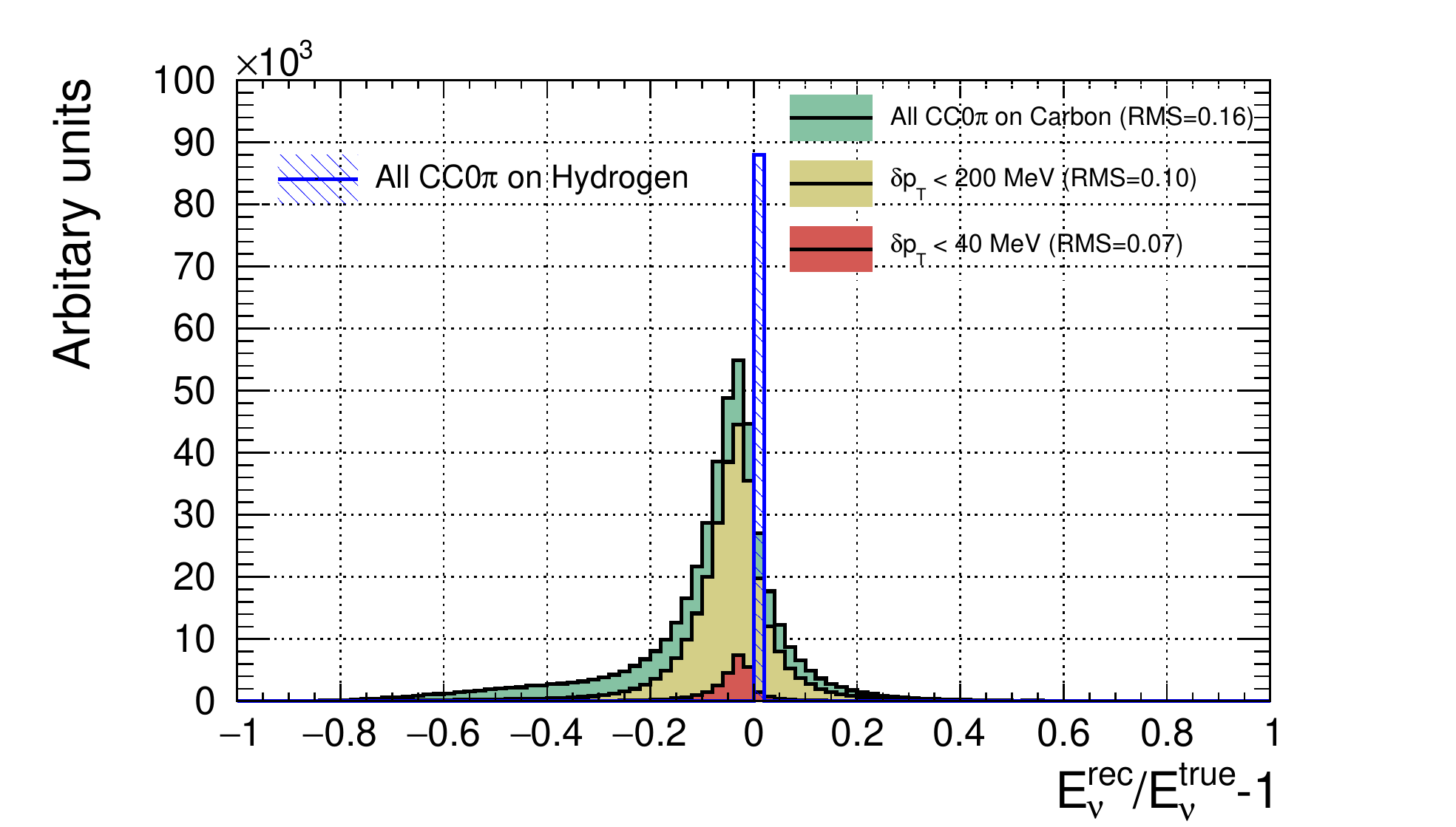}
\caption{\label{fig:trueStack} The bias and spread of the neutrino energy reconstruction for different regions of $\dpt$ for anti-neutrino interactions on carbon predicted by the NEUT 5.4.0 simulation. The neutrino energy is reconstructed using the kinematic method discussed in Sec.~\ref{sec:introduction} with no binding energy considered. The neutrino energy reconstructed perfectly from the hydrogen events is also shown, correctly normalised relative to the size of the carbon contribution. The systematic offset of the carbon contribution relative to the hydrogen is due to the nuclear binding energy in carbon interactions. The legend also shows the RMS as an indicator for the spread of the neutrino energy for each $\dpt$ region. No detector smearing or acceptance effects are considered.}
\end{figure}

The transverse momentum imbalance of anti-neutrino CC0$\pi$ interactions on commonly used hydrocarbon scintillator would be zero for the interactions on hydrogen and non-zero for those on carbon. This is demonstrated using events from the NEUT 5.4.0 neutrino interaction simulation~\cite{Hayato:2009zz, neutNewPaperInProgress} (discussed in Sec.~\ref{sec:simulation}) in Fig.~\ref{fig:dpt-no-smearing}, although in a real experiment the distribution accessible to the detector will be smeared and so a perfect isolation of hydrogen events is not feasible (this will be discussed in later sections). However, even the interactions on carbon nuclei with low $\dpt$ tend to be those which have been only mildly perturbed by FSI and 2p2h. For such an interaction to give low $\dpt$ the transverse momentum carried away by a second unseen nucleon (in 2p2h) or absorbed pion (from a pion absorption FSI) would have to cancel out the initial state imbalance of the observed nucleon, which is kinematically unlikely. Indeed, it can be seen from Fig.~\ref{fig:dpt-no-smearing} that there is a very low CCnonQE contribution at low $\dpt$. Fig.~\ref{fig:trueStack} additionally shows the bias and spread of the neutrino energy reconstruction (using the kinematic method discussed in Sec.~\ref{sec:introduction} assuming no binding energy) for different regions of $\dpt$, further demonstrating that the non-hydrogen interactions occupying the low $\dpt$ region are the ones in which the neutrino energy is better reconstructed. Overall it then follows that low $\dpt$ selection of events in an anti-neutrino beam may be able to give a heavily hydrogen enriched sample of events in which even the non-hydrogen interactions are also largely free of nuclear effects. These events can then be used to give a relatively unbiased neutrino energy reconstruction. 

Although some of the low-$\dpt$ CCQE events in carbon can be still affected by both the Fermi momentum and nuclear binding energy that can cause longitudinal imbalance ($\dplong$),  this corresponds to only a subset of them. This could  potentially be further mitigated using variables which additionally consider the inferred longitudinal imbalance of an interaction, such as proposed in~\cite{Furmanski:2016wqo}.

\subsection{Neutron detection}
\label{sec:neutron-detection-technique}

\begin{figure}
\centering
\includegraphics[width=9cm]{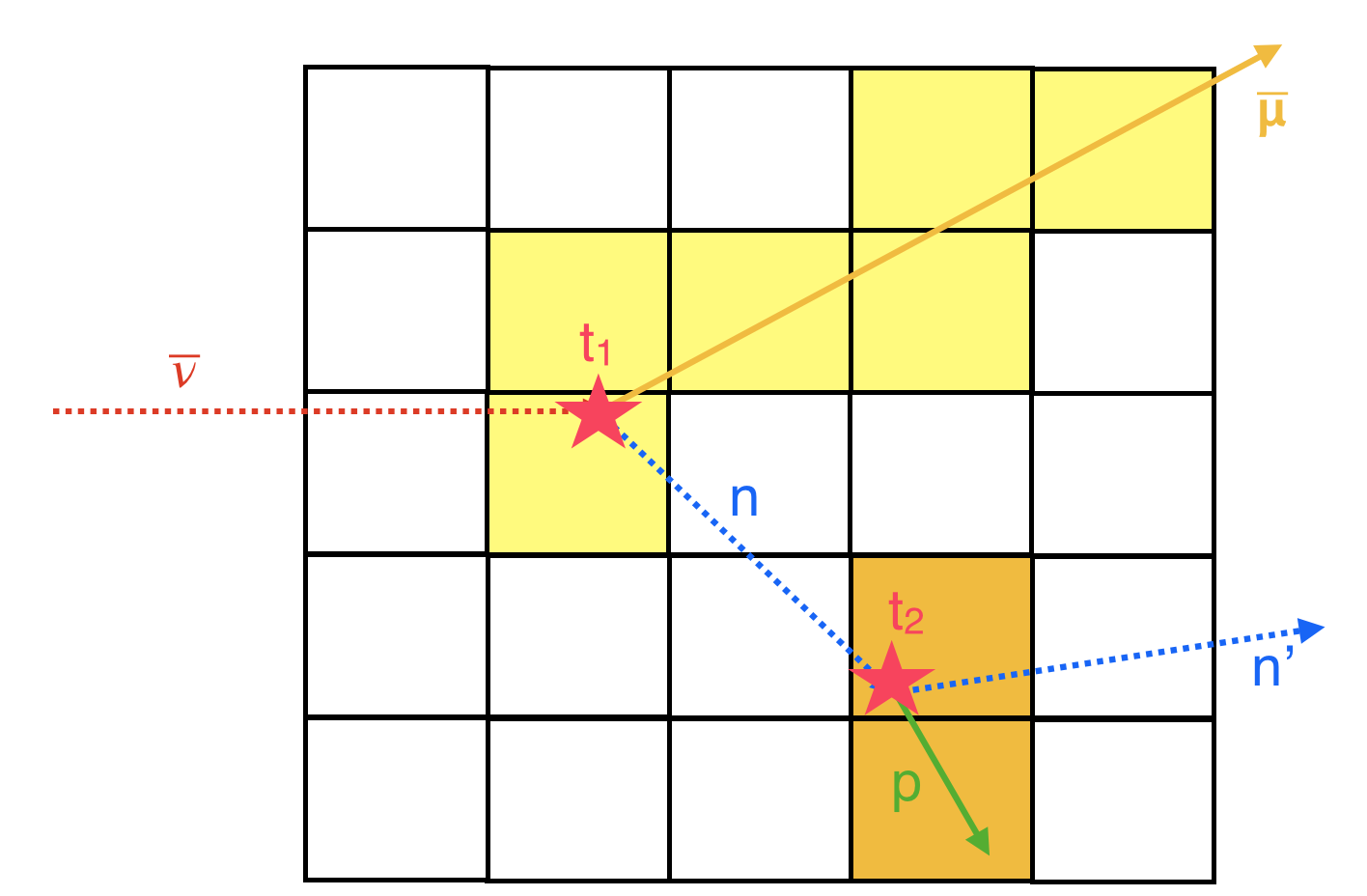}
\caption{\label{neutron-detection} Representation of an indirect neutron detection through the identification of a proton coming from a secondary neutron interaction. Each cube in the picture represents a detection element (e.g. a single scintillator cube) whilst the lines represent true particle trajectories. An anti-neutrino is shown entering the detector and interacting with a nucleus to produce a muon and a neutron at time $t_1$. The neutron then interacts at time $t_2$, ejecting a proton from a nucleus which is detected. The difference between $t_1$ and $t_2$ can be used to infer the neutron energy.}
\end{figure}

In order to measure $\dpt$ in anti-neutrino interactions, a precise detection and kinematic characterisation of neutrons is essential. In order to be able to measure their energy the neutrons must be detected before thermalisation (i.e. when they are still ``fast neutrons''). Fast neutrons must be directly identified through their scattering on protons or nuclei within an active detector medium. The charged secondary particles of these neutron interactions can then be identified as small, localised energy deposits within the detector. However, for neutrons of relevant kinematics (10 MeV - 1 GeV of kinetic energy) the amount of energy transferred to the secondary particles has only a very weak dependence on the initial neutron energy, so the neutron initial energy can only be inferred by measuring the Time-of-Flight (ToF) between the anti-neutrino vertex and the earliest ``neutron hit'', i.e. the measured energy deposited by the secondary particles. A schematic showing neutron detection by the measurement of the secondary particle recoil is shown in Fig.~\ref{neutron-detection}.

Such a measurement of fast-neutrons requires a detector with the following characteristics:
\begin{itemize}
    \item high neutron interaction probability within the active volume with detectable charged particles in the final state;
    \item 3-dimensional and 4$\pi$ tracking capability, to easily isolate the neutron hits and precisely measure the distance between the anti-neutrino vertex and the neutron interaction point; 
    \item very good time resolution, to provide a precise measurement of the neutron speed and the coincidence with the anti-neutrino vertex; 
    \item fully-active, to maximise the detection efficiency to the very first neutron interaction, as the recoiled particles of dead material often would not receive sufficient energy to produce a relatively long track in the detector;
    \item to provide a low-mass target for lower momentum fast neutrons the detector must  contain a significant mass of hydrogen nuclei. This is also required to identify anti-neutrino interactions free of nuclear effects.
\end{itemize}

An example of a detector technology that can fulfill all the above requirements is a plastic scintillator detector (containing predominantly $C_n H_n$ -like nuclei) with three readout planes. In addition to acting as an anti-neutrino interaction target, the presence of hydrogen provides a high interaction rate for neutrons below about 20~MeV kinetic energy, while the neutron - carbon cross section becomes dominant above this~\cite{Elkins:2019vmy}.
A recent analysis of fast neutron detection in neutrino interactions was performed by the MINERvA experiment, whose detector is based on the plastic-scintillator technology~\cite{Elkins:2019vmy}. However, the use of only two scintillator planes (so multiple hits are required for 3D tracking) and limited detector timing capability makes measurements of neutron kinematics in MINERvA extremely challenging.

In this manuscript we will focus on the detector technology proposed in~\cite{Sgalaberna:2017khy} as an active target in neutrino experiments. It consists of a fully-active 3D-granular plastic-scintillator detector made of 1~cm side optically isolated cubes. The scintillation light produced in each cube is read out by three orthogonal wavelength-shifting fibers along the X, Y and Z directions thereby allowing a single hit to provide a 3D position. 
Thanks to its geometry and a fast intrinsic time resolution of better than $0.6~\text{ns}$ for a Minimum Ionizing Particle (MIP) crossing a single cube~\cite{Mineev:2018ekk} (for a single fibre the timing resolution is $\sim0.95$ ns), it can satisfy all the requirements above. However the methodology described in this paper can be applied to any detector, or multi-detector system, with analogous characteristics.

\subsection{Simulation}
\label{sec:simulation}

We begin by using the T2K experimental neutrino flux~\cite{Abe:2012av,t2kfluxurl} and the NEUT 5.4.0 neutrino-nucleus interaction event generator~\cite{Hayato:2009zz, neutNewPaperInProgress} to study the viability of identifying such a sample of interactions quasi-free of nuclear effects in a realistic detector. Events are produced on a CH target before an experimentally accessible sub-sample of CC0$\pi$ interactions is selected. The version of NEUT 5.4.0 used simulates CCQE and 2p2h interactions according to the model of Nieves et. al.~\cite{Nieves:2011pp,Nieves:2011yp}, RES interactions using the model of Rein and Sehgal~\cite{Rein:1981ys} and a semi-classical cascade model to describe FSI~\cite{Hayato:2009zz}.

Separately, the response of the detector to fast neutrons was studied with isotropic ``particle guns'' using the GEANT 4.10.5 simulation software~\cite{Agostinelli:2002hh}. All relevant detector response effects, such as the light quenching in the plastic, the light capture efficiency and light attenuation in the fiber and the Silicon PhotoMultiplier (SiPM) photo-detection efficiencies, are taken into account. The output of the detector response is given in terms of SiPM photoelectrons (PE). The neutron hits are clustered together (into a ``neutron cluster'') and a cut to omit all hits within a $3\times3\times3$ cube volume around the neutron production point is applied to account for difficulties in separating neutron-interaction induced hits from those coming from the primary anti-neutrino interaction. 

\begin{figure}
\centering
\includegraphics[width=8cm]{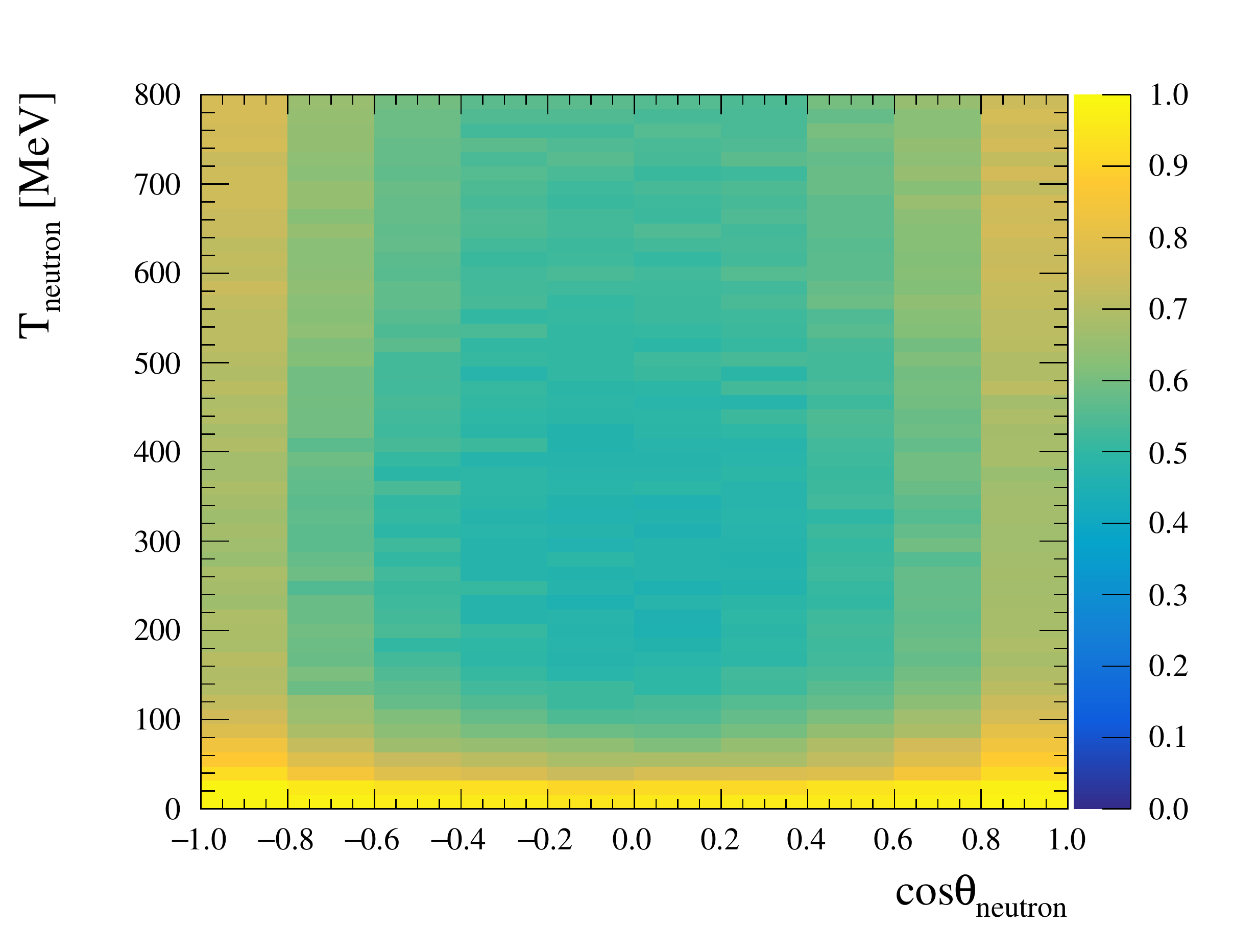}

\caption{\label{fig:neutron-eff} The neutron detection efficiency for the detector proposed in~\cite{Sgalaberna:2017khy} ($2\times0.6\times2$~m$^3$) determined from particle gun simulations. The detector chosen for the rest of this manuscript has a size of $2\times2\times2$~m$^3$ and so has an approximate isotropic efficiency with a momentum dependence similar to the first/last  $\cos{\theta_{neutron}}$ bin in this figure.}
\end{figure}

The neutron detection efficiency is closely tied to the size of the simulated detector, as a larger volume means that more neutron interactions are contained and identified. Fig.~\ref{fig:neutron-eff} shows the neutron detection efficiency, established using the aforementioned particle guns, for a detector the size of the one proposed in~\cite{Sgalaberna:2017khy} ($2\times0.6\times2$ m$^3$), clearly showing a lower efficiency for the high angle neutrons that are less likely to be contained. For generality and simplicity, we consider a $2\times2\times2$ m$^3$ detector for the rest of this manuscript (where the detector efficiency is approximately equivalent to the first/last $\cos{\theta_{neutron}}$ bin in Fig.~\ref{fig:neutron-eff} and is isotropic in angle). Combining the $2\times2\times2$ m$^3$ detector efficiency with NEUT's prediction of outgoing neutron kinematics for CC0$\pi$ interactions, an integrated neutron detection efficiency of 71\% is found (this is reduced to 50\% for the $2\times0.6\times2$ m$^3$ detector).

The primary neutron candidate consists of the first detected cluster of hits after the anti-neutrino vertex time.  
As a first order approximation, the time resolution of the single neutron cluster is obtained by:
\begin{equation}
    \sigma_{t}^{ly} = \left \{ 0.95~\text{ns} / \sqrt{3} \right \} 
                \cdot \sqrt{ 40~\text{PE} / \text{LY} }, 
    \hspace{2mm}
    \sigma_{t}^{ly} > 200~\text{ps}
\label{eq:time-res-opt}
\end{equation}
where 0.95~ns and 40~PE are respectively the approximate time resolution (as reported in~\cite{Mineev:2018ekk}) and light yield of a single readout channel for a MIP crossing one cube, LY is the light yield of the neutron cluster, i.e. the total number of detected PE. The normalisation by $\sqrt{3}$ is applied because three WLS fibers collect scintillation light from a single cube. Here only the prompt light yield is considered (which is produced within 200~ps of the neutron interaction) in order to avoid an artificially enhanced timing resolution due to additional neutron re-scattering or tertiary particle interactions.

A more conservative timing resolution formula, e.g. in case the electronics speed provides limits when more light than a MIP is produced, can be given by: 
\begin{equation}
    \sigma_{ch}^{ly} = \left \{ 0.95~\text{ns} / \sqrt{\# \text{channels} } \right \}, 
    \hspace{2mm}
    \sigma_{t}^{ch} > 200~\text{ps}
\label{eq:time-res-cons}
\end{equation}

where $\# \text{channels}$ is the number of readout channels that measured more than one PE and which were triggered within 200~ps of the neutron interaction. 
Simulation results will be shown by computing the neutron ToF
with both Eq.~\ref{eq:time-res-opt} and~\ref{eq:time-res-cons}.
Both equations limit the timing resolution of neutron candidate to be worse than 200~ps to effectively simulate an arbitrary limitation due to some front-end electronics sampling rate. 

As the timing resolution improves if more scintillation light is produced or more readout channels are triggered, a quality cut is applied on the minimum number of detected PEs in the neutron hit cluster. In order to select only events with a timing resolution at least as good as the one for a detected MIP, the event is rejected if $\text{LY} < 40~\text{PE}$. The time resolution of the interaction vertex was neglected, assuming that it will not be a limiting factor as it can be measured by backward extrapolating in time the $\mu^{+}$ track. 

\begin{figure}
\centering
\includegraphics[width=9cm]{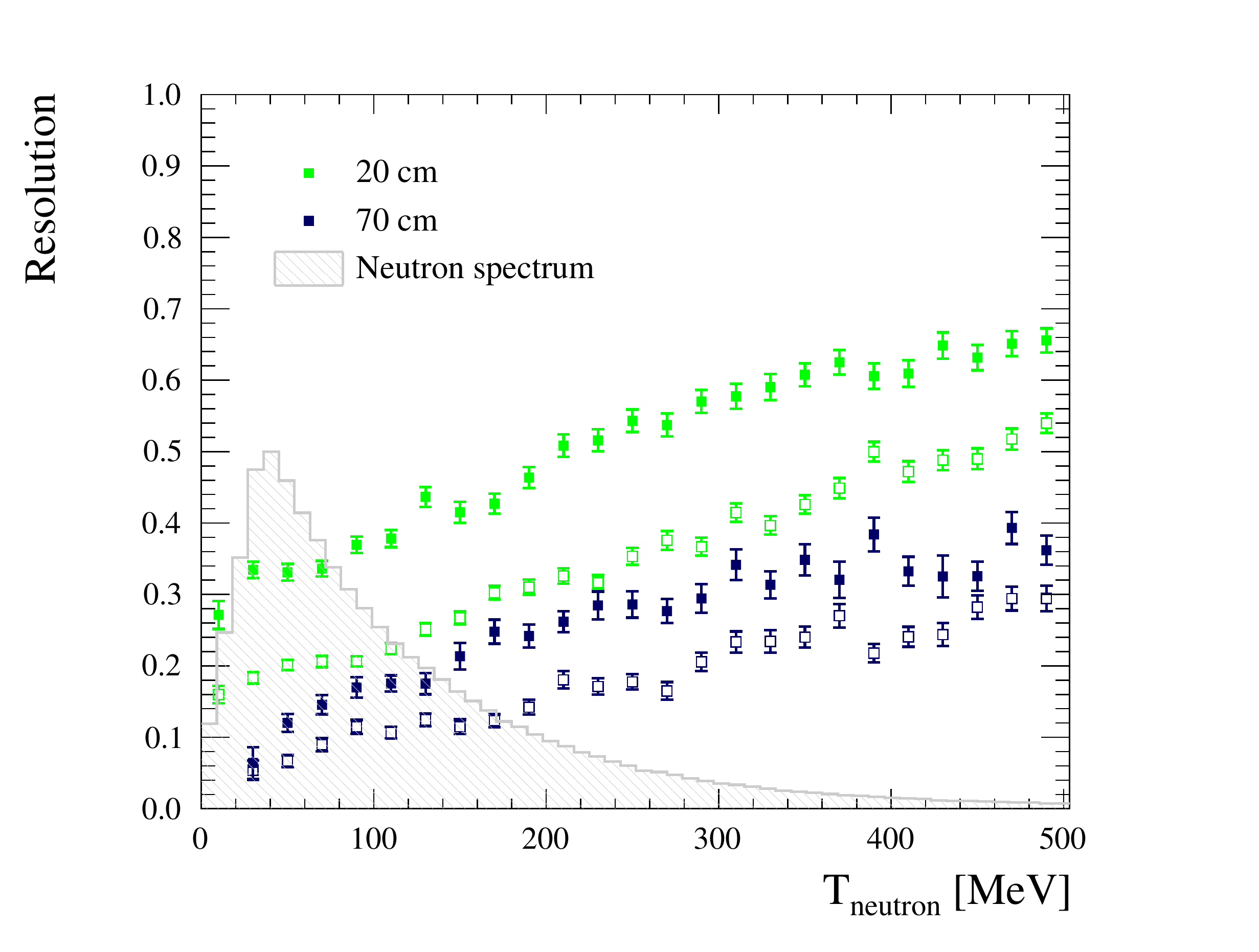}

\caption{\label{fig:neutron-resolution-leverarm} The neutron kinetic energy resolution as a function of the neutron's kinetic energy, assuming different lever arm cuts and timing resolutions. The hollow markers correspond to a timing resolution using Eq.~\ref{eq:time-res-opt} whilst the filled markers correspond to Eq.~\ref{eq:time-res-cons}. The resolution is taken as the ratio of the largest of the two standard deviations to the mean of a double-sided Gaussian fitted to the reconstructed neutron kinetic energy in a bin of true neutron kinematic energy. The grey band shows NEUT 5.4.0's predicted distribution of neutron kinematic energies for CC0$\pi$ neutrino interactions using the T2K experiment's anti-neutrino flux.
}
\end{figure}

Furthermore, additionally requiring some minimum distance between the anti-neutrino vertex and the neutron cluster (``lever arm'') provides a sample of neutrons with an improved ToF reconstruction and, consequently, a better neutron energy resolution. Fig.~\ref{fig:neutron-resolution-leverarm} shows how the neutron kinetic energy resolution varies as a function of the neutron's kinetic energy, assuming different lever arm cuts and timing resolutions. It can be seen that if a cut of 70~cm on the neutron lever arm is applied and Eq.~\ref{eq:time-res-opt} is used to estimate the resolution on the neutron ToF, the neutron kinetic energy can be measured with a resolution between 10-20\% for energies up to about 300 MeV. Improving the time resolution with respect to the conservative assumption made in Eq.~\ref{eq:time-res-opt} would further improve the neutron energy resolution.

\subsection{Analysis strategy}
\label{sec:analysis-strategy}

A sample of experimentally accessible CC0$\pi$ events (i.e. those which have $p_{\mu^{+}}>$ 100 MeV/c) is first selected from the events generated using NEUT 5.4.0. Events with protons in the final state are not rejected but the proton kinematics are not used to compute the reconstructed anti-neutrino energy and event $\dpt$. This approach is conservative as a 3D fine grained detector should be able to reconstruct protons with momenta both higher than 300~MeV/c \cite{Sgalaberna:2017khy} and lower by detecting at the vertex an amount of scintillation light higher than the one expected for a single $\mu^+$. The assumption that CC0$\pi$ events are experimentally accessible is justified by the fact that selecting this type of final state is relatively easy and previous analyses looking at this channel have been high in purity ($\sim 80\%$)~\cite{Abe:2018pwo}. In a 3D fine grained detector the situation is even better as it is very easy to identify a single MIP-like track without any vertex or deposited energy around its starting point to create an extremely pure CC0$\pi$ sample. Moreover, the small background is not expected to contribute preferentially to the interesting region of low transverse kinematic imbalance. 

The kinematics of the outgoing neutron from these events is then smeared and the detection efficiency corrected according to the results of the GEANT4 simulation, giving an effective simulation of the kinematics that would be made in a full detector simulation chain. Additionally, an effective travel distance was assigned to each neutron depending on its true kinetic energy, as evaluated with the GEANT4 simulation.

Several different smearings are applied to allow different choices of the detector timing resolution and the chosen lever arm cut. The momentum of the outgoing $\mu^+$ is then smeared by 4\% using a Gaussian distribution, conservatively driven by the typical resolution of a spectrometer detecting muons that escape from the scintillator, whilst the azimuthal and polar angles are smeared by 1$^{\circ}$ (chosen based on the 1 cm granularity of the detector).

Once the reconstructed (smeared) kinematical variables are computed, the smeared momentum imbalance on the plane transverse to the incoming neutrino is computed for each event as in Eq.~\ref{eq:dpt} (using only the highest momentum neutron if there is more than one). 

The anti-neutrino energy is reconstructed for each event using the kinematic method described in Sec.~\ref{sec:introduction} as (no binding energy correction is applied):
\begin{equation}
\label{eq:enu}
    E_{\nu} = \frac{m_n^2 - m_p^2 - m_{\mu}^2 + 2m_{p}E_{\mu}}{2(m_p-E_{\mu}+p_{\mu}\cos{\theta_\mu})},
\end{equation}
where $m_n$, $m_p$ and $m_\mu$ are the masses of a neutron, proton and muon respectively whilst $E_\mu$, $p_\mu$ and $\cos{\theta_\mu}$ are the energy, momentum and angle of the outgoing $\mu^+$ with respect to the incoming neutrino. The smeared lepton kinematics are used such that the reconstruction of neutrino energy includes detector resolution effects.

The strategy described above was chosen in order to be able to quickly switch between different anti-neutrino interaction models implemented in NEUT 5.4.0 without rerunning the entire simulation chain. However, all the relevant detection smearing effects relevant for this analysis have been carefully evaluated with GEANT4 and included in this fast simulation.

\section{Results}
\label{sec:results}

\begin{figure}
\centering
\includegraphics[width=9cm]{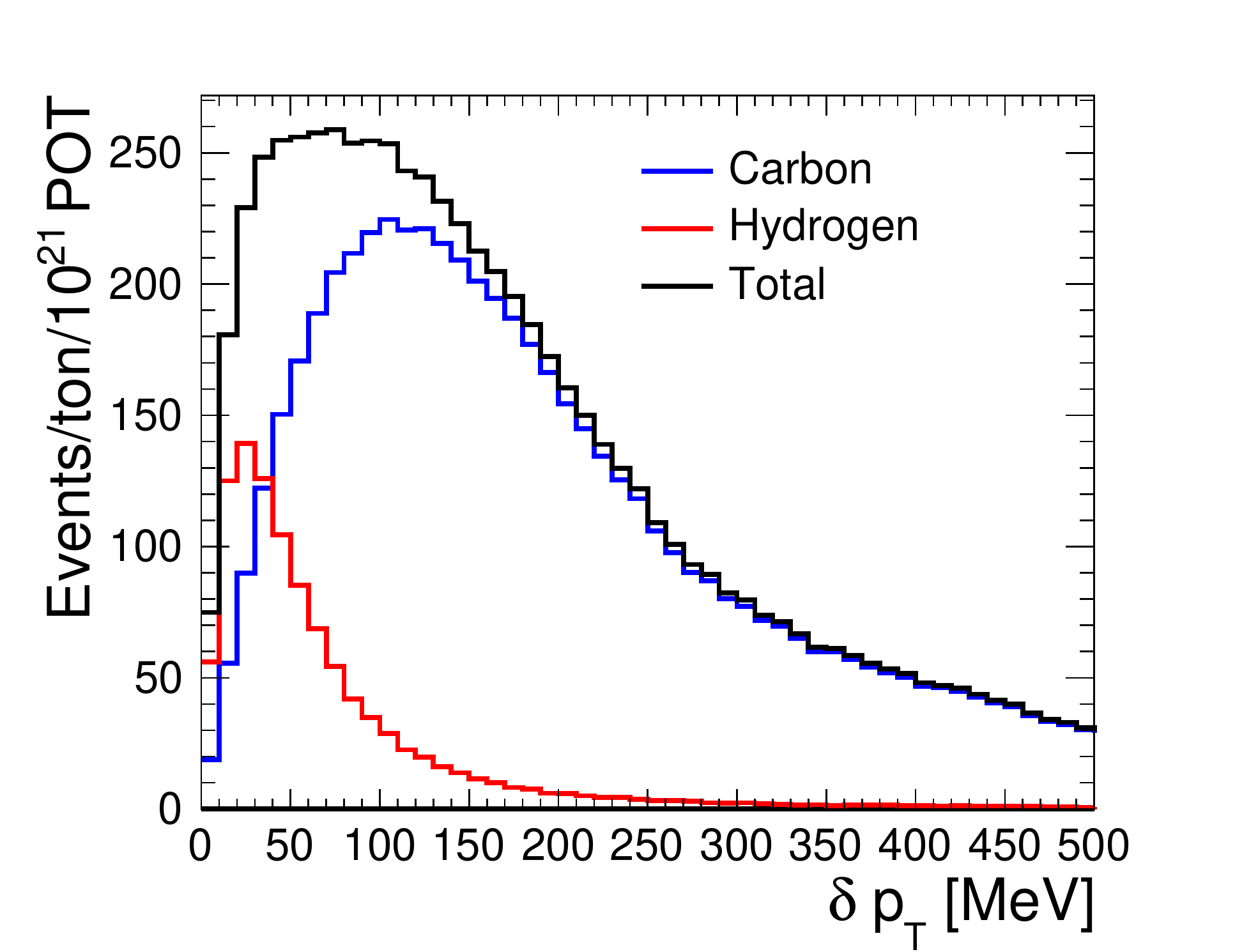}
\caption{\label{fig:dpt-smearing} The NEUT 5.4.0 predicted event rate of CC0$\pi$ interactions from the T2K anti-neutrino flux as a function of $\dpt$ obtained after applying the detector smearing effects as described in Sec.\ref{sec:analysis-strategy} with a 10 cm lever-arm cut and using a timing resolution given by Eq.~\ref{eq:time-res-opt}. Events are separated based on the target nucleus of the neutrino interactions.}
\end{figure}

The $\dpt$ distributions calculated using realistic detector smearing as described in Sec.~\ref{sec:analysis-strategy} is shown in Fig.~\ref{fig:dpt-smearing} (for a 10 cm lever-arm cut). A comparison of Figs.~\ref{fig:dpt-no-smearing} and~\ref{fig:dpt-smearing} shows that the visible hydrogen contribution at low $\dpt$ remains largely distinct from the carbon contribution after the application of detector effects. As described in Sec.~\ref{sec:dpt}, we then attempt to obtain a sample of events influenced minimally by nuclear effects by requiring $\dpt$ to be below a certain value. 

\begin{figure}
\centering
\includegraphics[width=10cm]{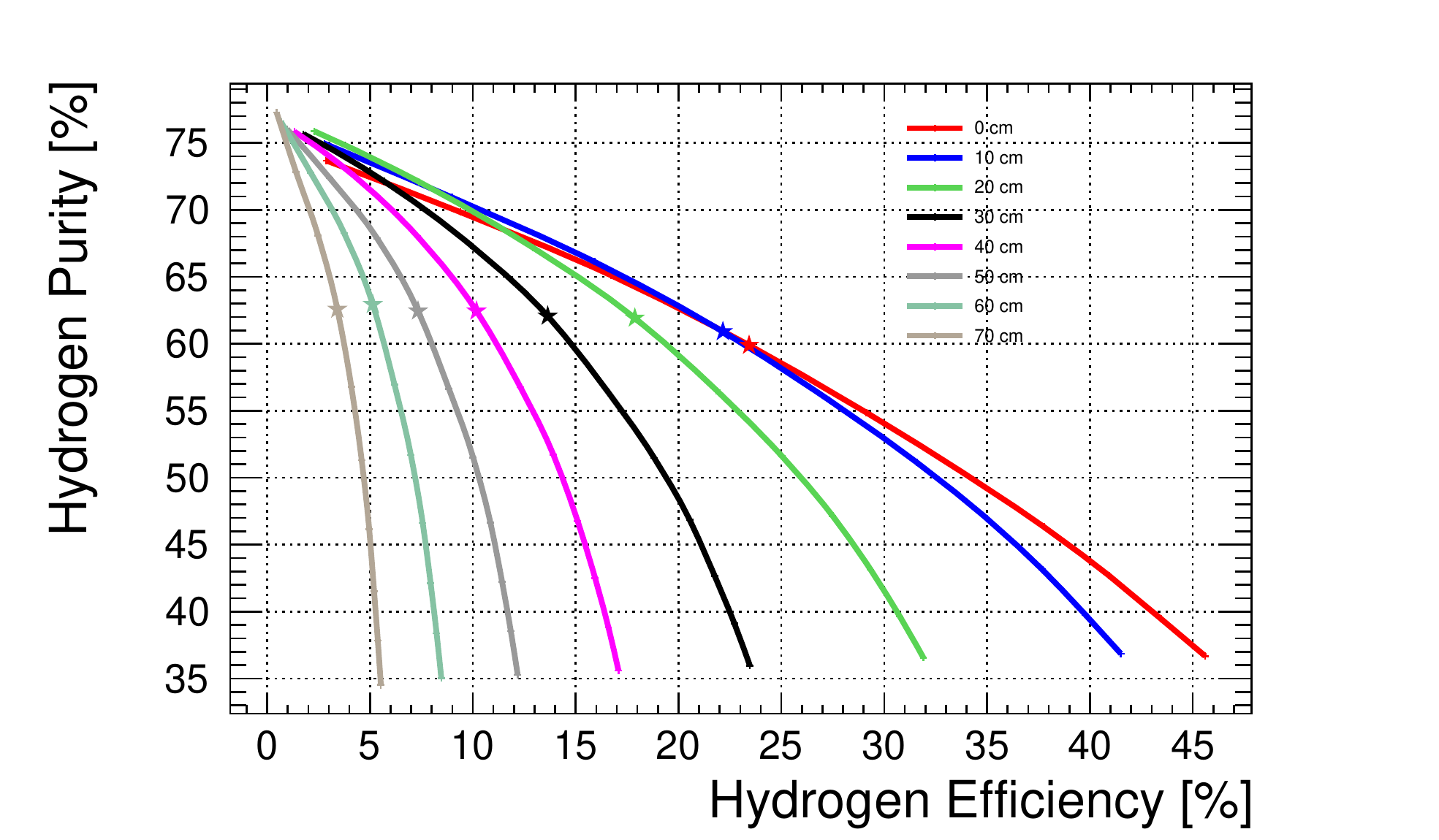}
\caption{\label{fig:purEff} The anti-neutrino hydrogen purity vs efficiency for different $\dpt$ and lever-arm cuts. The first (top left) marker on each line corresponds to a 10~MeV cut and then the star corresponds to the chosen $\dpt$ cut of 40~cm. Each line corresponds to a different lever-arm cut and are made using Eq.~\ref{eq:time-res-opt} to determine the time resolution.}
\end{figure}

The cut on $\dpt$ and on the lever-arm is chosen to obtain a high purity of anti-neutrino hydrogen events whilst maintaining a reasonable sample size. This optimisation is shown in Fig.~\ref{fig:purEff} for the timing resolution given by Eq.~\ref{eq:time-res-opt}. If Eq.~\ref{eq:time-res-cons} is used instead the lines follow a similar shape but the purity is approximately 10\% smaller for the same efficiency. Overall a good compromise is achieved for $\dpt < 40~\text{MeV/c}$ and a 10~cm lever-arm cut, allowing a hydrogen purity and efficiency of around 61\% and 22\% respectively, corresponding to approximately 988 anti-neutrino interactions with either hydrogen or carbon per ton per $10^{21}$ POT (assuming Eq.~\ref{eq:time-res-opt}).

Fig.~\ref{fig:purEff} demonstrates that a stronger lever arm cut worsens the efficiency to select anti-neutrino hydrogen events. However, differently from what one may expect, it does not always improve the purity despite improving the neutron energy resolution (as shown in Fig.~\ref{fig:neutron-resolution-leverarm}). This is because there is some correlation between the neutron lever arm and its initial energy, i.e. lower energy neutrons in general travel a shorter distance. Therefore, choosing a longer lever arm selects faster neutrons and so they do not necessarily have a much greater ToF. Furthermore, neutrons coming from neutrino hydrogen interactions tend to be of a slightly higher energy (as they are not slowed by nuclear binding energy or FSI). More details can be found in Appendix~\ref{app:leverarm}.

The selection cuts can considerably change depending on the experimental conditions. In an experiment dominated by the low statistics the event selection may favor a high anti-neutrino hydrogen selection efficiency compared to the purity, while in a high-statistics experiment the purity may be maximised with a reduced efficiency. For example, by requiring $\dpt < 20~\text{MeV/c}$ a purity of almost 75\% could be achieved. Although Fig.~\ref{fig:purEff} demonstrates that the selection efficiency is largely driven by the lever arm and $\dpt$ cuts, the containment of neutrons inside the $2\times2\times2$ m$^3$ detector volume is also important. As discussed in Sec.~\ref{sec:neutron-detection-technique}, only 71\% of neutrons are contained, representing an approximate upper limit on the hydrogen efficiency. A larger detector would therefore have the potential to further increase the efficiency.

\begin{figure}
\centering
\includegraphics[width=9cm]{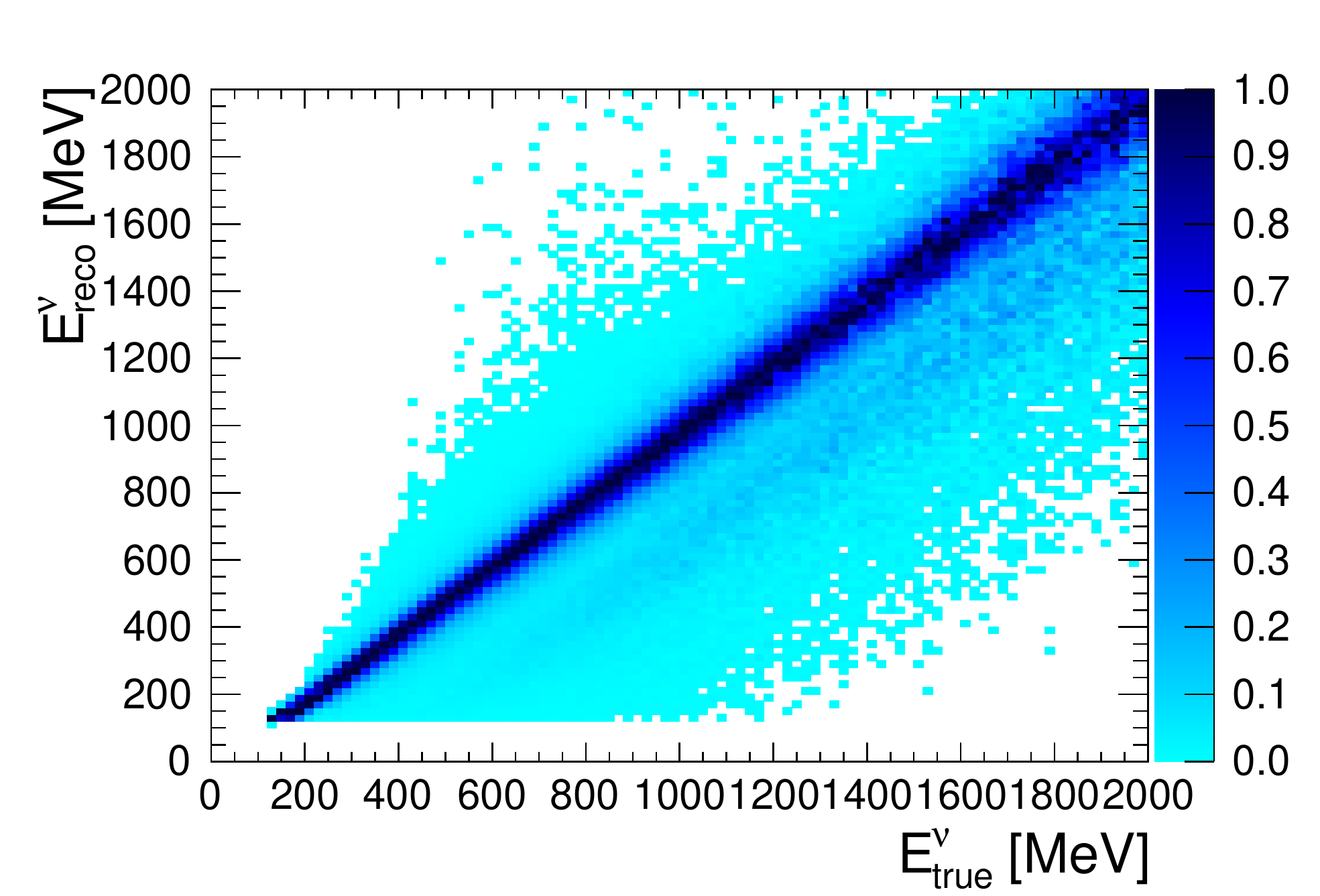}
\includegraphics[width=9cm]{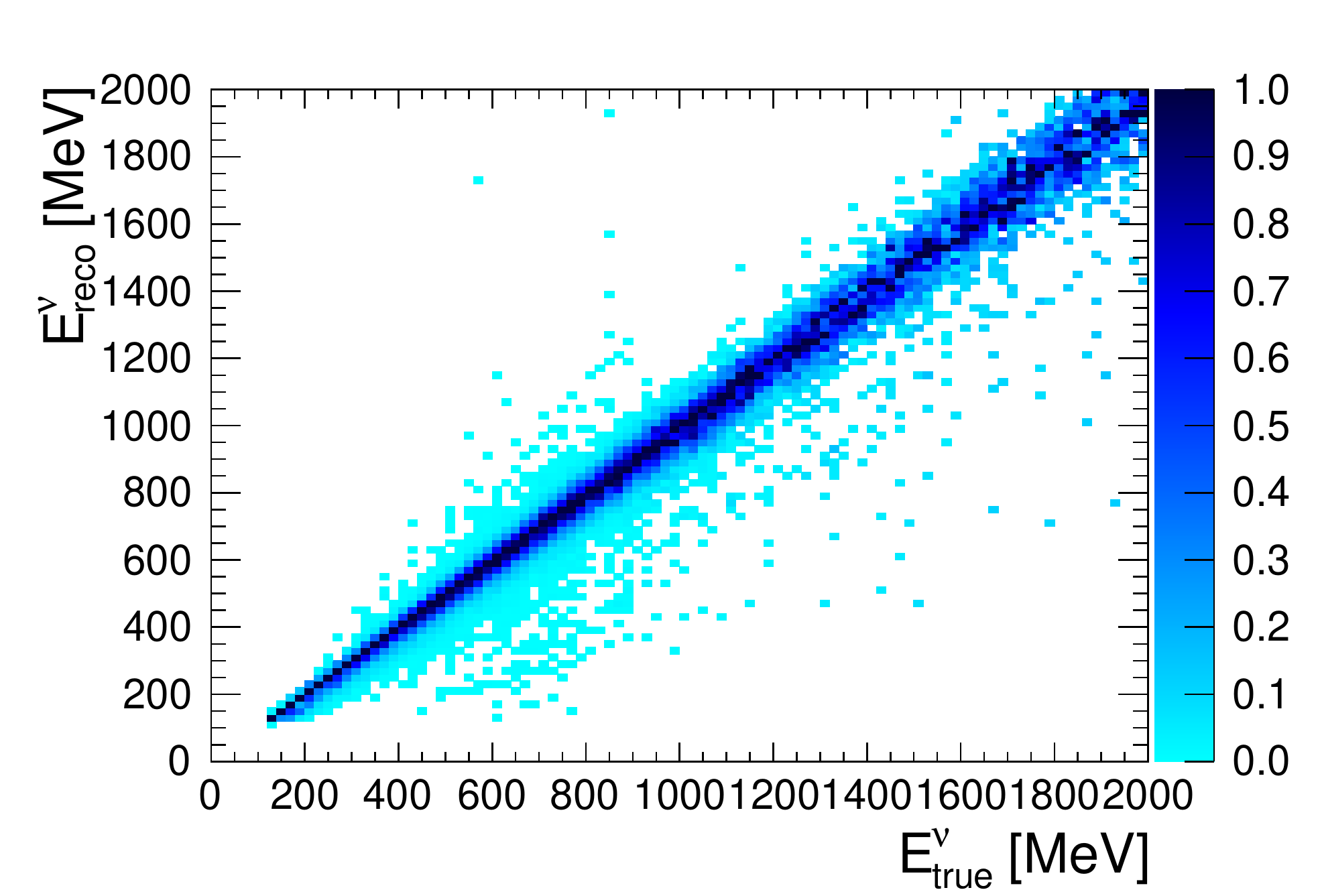}
\caption{\label{fig:neutrino-energy-2D} The mapping from true to reconstructed neutrino energy including smearing from both the nuclear effects described in Sec.~\ref{sec:introduction} and the detector effects described in Sec.\ref{sec:analysis-strategy} before applying the $\dpt$ and lever arm cuts (above) and after (below). The timing resolution used to build $\dpt$ is the one given by Eq.~\ref{eq:time-res-opt}. The z-axis of the histograms is normalised such that the largest value in each column is one.
}
\end{figure}

The events passing the $\dpt$ and lever arm cuts are then used to reconstruct anti-neutrino energy using Eq.~\ref{eq:enu} and the subsequent resolution on the anti-neutrino reconstructed energy is shown in Fig.~\ref{fig:neutrino-energy-2D}, where the cut is shown to improve the resolution from 15\% to around 7\% using either Eq.~\ref{eq:time-res-opt} or~\ref{eq:time-res-cons} for the timing resolution (the change in the timing resolution alters the number of events passing the cut more than the resolution of events which pass it). The contribution of different interaction modes to the selected events are further analysed in Appendix~\ref{app:selectedevents}.

\begin{figure}
\centering
\includegraphics[width=10cm]{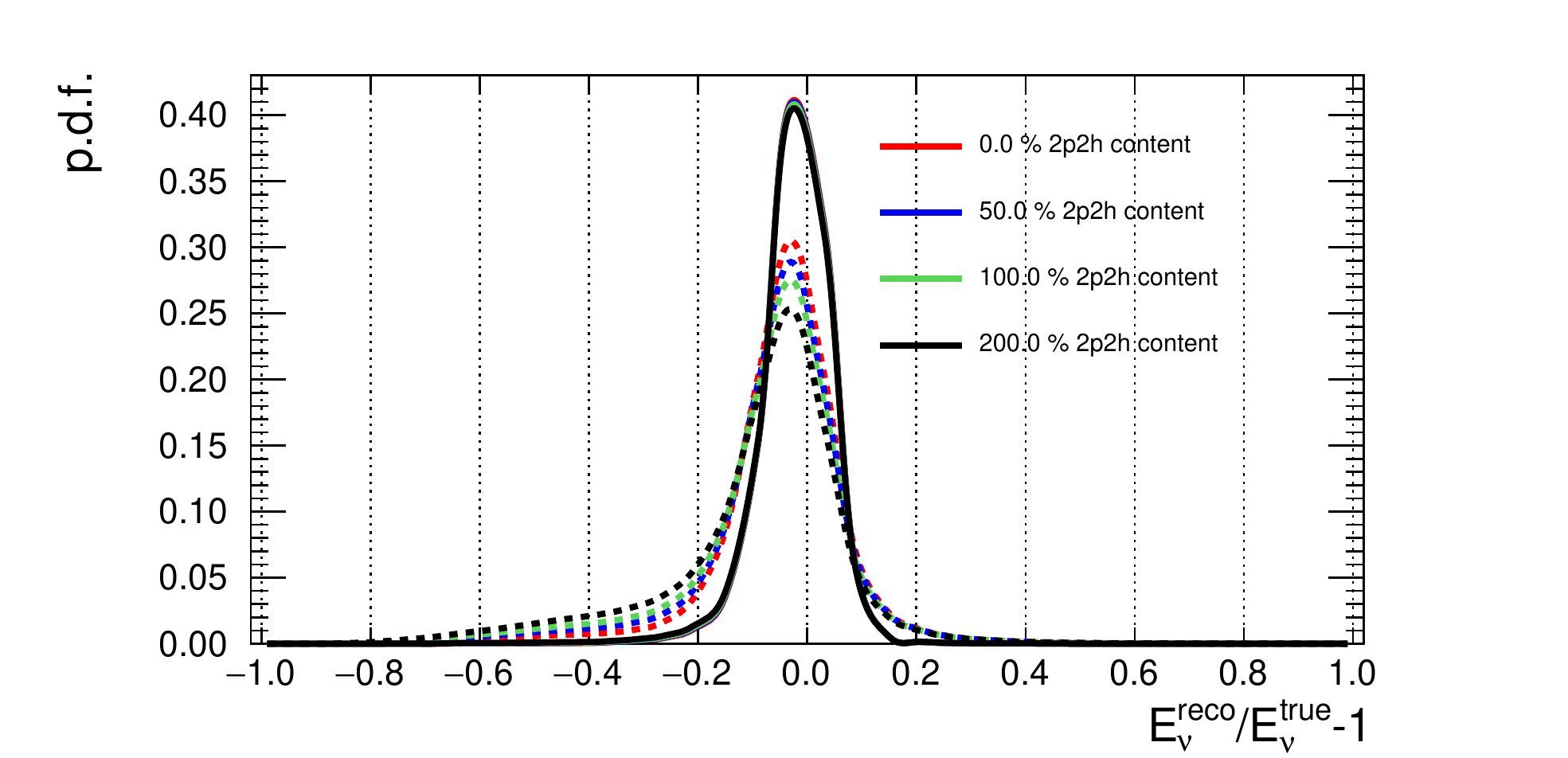}
\caption{\label{fig:neutrino-energy-2p2h} The anti-neutrino energy resolution for a minimum neutron lever arm of 10~cm is shown. Eq.~\ref{eq:time-res-opt} was used to compute the time resolution. Solid lines shows the energy resolution after applying the cut on $\dpt$ (10 cm and 50 MeV respectively), while dashed lines were obtained without any cut on the momentum imbalance. The line colours correspond to different normalisations of the $2p2h$ component. 
}
\end{figure}

\begin{figure}
\centering
\includegraphics[width=10cm]{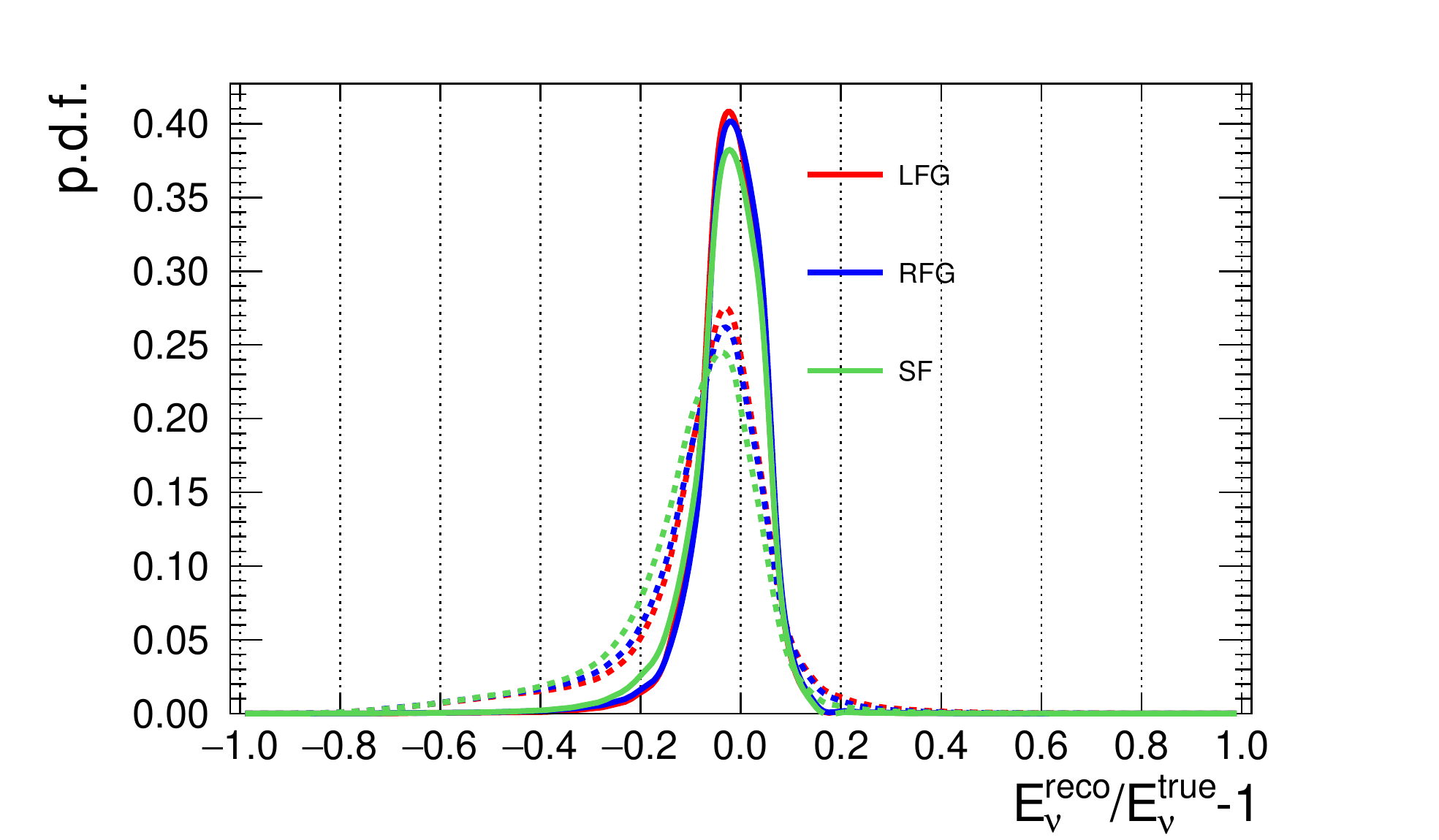}
\caption{\label{fig:neutrino-energy-pf}  The anti-neutrino energy resolution for a minimum neutron lever arm of 10~cm is shown. Eq.~\ref{eq:time-res-opt} was used to compute the time resolution. Solid lines shows the energy resolution after applying the cut on $\dpt$, while dashed lines were obtained without any cut on the momentum imbalance. The line colors correspond to different models of the initial nucleon's momentum and binding energy in CCQE interactions. LFG is the local fermi gas of Nieves and collaborators~\cite{Nieves:2011pp,Nieves:2011yp}, RFG is the relativistic fermi gas model of Smith and Moniz~\cite{Smith:1972xh} and SF is the Spectral Function model of Benhar and collaborators~\cite{Benhar:1994hw}.
}
\end{figure}

As discussed in Sec.~\ref{sec:dpt}, the spread of the neutrino energy for the low $\dpt$ events should be largely controlled simply by the detectors ability to reconstruct the outgoing lepton kinematics and not so much by poorly understood nuclear effects. To demonstrate this, the reconstruction of the anti-neutrino energy for different normalisations of the $2p2h$ interaction component is analysed in Fig.~\ref{fig:neutrino-energy-2p2h} and for different models of the carbon nuclear ground state in Fig.~\ref{fig:neutrino-energy-pf}.

\begin{figure}
\centering
\includegraphics[width=9cm]{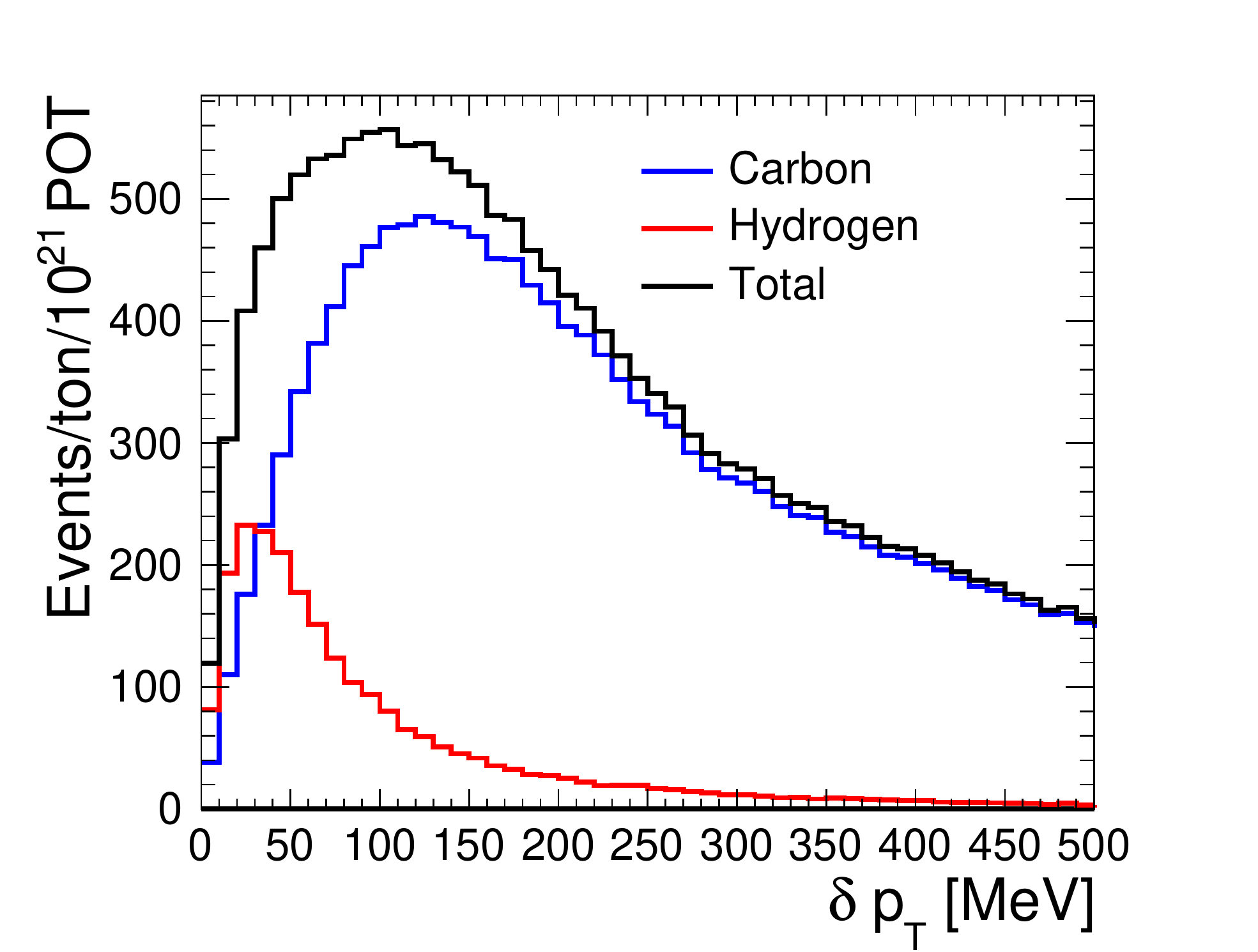}

\caption{\label{fig:neutrino-energy-dune-flux} The equivalent of Fig.~\ref{fig:dpt-smearing} using the DUNE neutrino energy flux, a 10 cm lever-arm cut and the timing resolution given by Eq.~\ref{eq:time-res-opt}. Events are separated based on the target nucleus of the neutrino interactions.  
}
\end{figure}

The same analysis was also performed using the anti-neutrino on-axis flux predicted for the DUNE experiment~\cite{dune-beam-1,dune-beam-2,dunefluxurl}, where the energy peak is between 2 and 3 GeV, and shown to give analogous improvements in anti-neutrino energy reconstruction: using the calorimetric method discussed in Sec~\ref{sec:introduction} the resolution improves from $\sim$10\% to $\sim$5\% for a 40 MeV cut on $\dpt$ and a 10 cm lever-arm cut. The predicted event rate of CC0$\pi$ interactions from the DUNE anti-neutrino flux as a function of $\dpt$, obtained after applying the detector smearing effects, is shown in Fig.~\ref{fig:neutrino-energy-dune-flux}. In general neutrons from interactions at DUNE energies travel faster and so resolution to their energy is slightly worse, however having such an intense beam at a somewhat higher energy may allow this to be mitigated whilst maintaining a reasonable number of events by requiring larger lever arm and smaller $\dpt$ cuts. 

\section{Discussion}
\label{discussion}

Overall it is clear that, under the conditions of the simulations presented in Sec.~\ref{sec:methodology}, the demonstrated method of anti-neutrino energy reconstruction is able to substantially improve the resolution and mitigate the impact of biasing nuclear effects with respect to traditional methods. Turning the measured event rate into a flux constraint requires knowledge of the anti-neutrino nucleon cross-section, which is known to a level of 5\% or better within the relevant kinematic phase-space~\cite{Megias:2019qdv}. The method also relies on a good understanding of neutron interaction cross sections, which are currently known to the level of 1\%-3\%~\cite{BOWEN1961640} at relevant energies, and of the detector response to the protons coming from the neutron interactions, which is expected to be able to be controlled with test-beam data and suitable control samples. Therefore, neglecting the uncertainties on the carbon contribution to the events passing the $\dpt$ and lever-arm cuts which are discussed below, it is expected that a flux constraint with a precision of 5\% or better could be achieved. 

As predicted in Sec.~\ref{sec:dpt}, the $\dpt$ cut on the carbon interactions is particularly effective at mitigating biases from interactions with multi-particle initial states, as shown in Fig.~\ref{fig:neutrino-energy-2p2h}, which offers more modest improvements when considering variations of the nuclear ground state for carbon CCQE interactions, as shown in Fig.~\ref{fig:neutrino-energy-pf}. It therefore remains critical to understand the size and spread of the carbon CCQE interactions (making up $\sim 39\%$ of events passing the $\dpt$ and lever-arm cuts) at a level of precision which does not undermine the flux constraint. It is hoped that this may be achieved by further analysing \textit{neutrino} interactions at low $\dpt$ (where $\dpt$ is formed from the outgoing muon and proton), as has already been demonstrated in references~\cite{Abe:2018pwo,Lu:2018stk}, albeit with some limitations related to proton tracking thresholds. A future detector with the qualities specified in Sec.~\ref{sec:neutron-detection-technique} would be expected to overcome these limitations (see, for example, Sec.~6.5 of~\cite{Abe:2019whr}) with a much larger sample of neutrino interactions than it would have anti-neutrino due to their higher cross-section (for the T2K beam around 3.3 times more CC0$\pi$ neutrino interactions are expected for the same exposure).

It should also be noted that this analysis conservatively does not eliminate interactions with protons in the final state, which are likely to be those in which the neutrino energy will not be well reconstructed or where the neutron has undergone FSI. A rejection of these events may therefore further improve neutrino energy resolution. This would in principle be possible by looking for energy deposits close to the neutrino interaction vertex larger than what would be expected from the observed muon in a scintillator detector. An alternative method of rejecting such protons would be to use an extremely precise tracking detector, such as a gaseous time projection chamber (TPC), with a few MeV kinetic energy detection threshold, combined with supplementary detectors to ensure efficient neutron detection capability.

Beyond rejecting events with protons, several other extensions to the analysis method are also possible. For example, although this analysis was only focused on pion-less interactions, the same method can also be used for 3-particle CC-1$\pi$ final states, where $\dpt$ is constructed considering the kinematics of all the outgoing particles. In particular, the same analysis method can be applied to $\bar{\nu}_{\mu} + p \rightarrow \mu^+ + n + \pi^0$ events, where a sample of CC$\pi^0$ candidate events is obtained by selecting an additional isolated $\pi^0$. Other channels with protons and charged pions in the final state are possible, but proton detection thresholds tend to be quite large relative to their typical kinematics limiting the validity of the measurement to interactions with suitably high momentum transfer. Moreover, accurate three track reconstruction is extremely challenging unless all tracks are fairly long, again limiting the applicability of this approach.

% Improvement in the detector technology
The proposed analysis method was also only applied to commonly used polystyrene-based plastic scintillator where the chemical composition is $(\textit{C}_8 \textit{H}_8)_n$. However, if a detector could be made with the same basic detection properties, but a higher concentration of hydrogen, the purity and number of hydrogen interactions clearly increases. For example, mineral oil contains about two hydrogen nuclei per carbon nucleus. %If the same analysis method can be performed with exactly the same performances, the rejection of anti-neutrino - carbon interactions would increase from 68\% to 81\%.

A potential advantage of this method with respect to other flux-constraining techniques is that it can extract information on both the normalisation and the shape of the anti-neutrino flux. Other methods based on the precise detection of the (anti)neutrino-electron scattering are mainly sensitive to the flux normalisation but less to its shape and cannot distinguish between neutrinos and anti-neutrinos. Moreover, the neutrino-electron cross section becomes quite low at energies below 1 GeV, where the method proposed here is most relevant. It is also interesting to consider combining this method with ``low-$\nu$'' based flux constraints as described in~\cite{paper-low-nu-1,paper-low-nu-2,paper-low-nu-3}.  A detector that can perform complementary measurements of the (anti)neutrino flux would allow the strongest and best validated flux constraint. Furthermore, if another flux constraint was to provide a much more accurate constraint on the flux normalisation, the method presented here can use this to place a powerful constraint on the nuclear form factors. 

The method presented here is also complimentary to those that use other observables to isolate interactions on hydrogen from pion production interactions, such as $\dptt$ as proposed in~\cite{Lu:2015hea,Duyang:2019prb}. These variables can only be used for events with more than two final state particles (pre-FSI) and so are more sensitive to the flux only at neutrino energies above $~1$ GeV (where pion production becomes prominent) whilst the method presented here excels for lower energy CCQE interactions, that span the neutrino energy ranging from the sub-GeV to the multi-GeV region. This is especially important for experiments like T2K/Hyper-K with a fairly low energy neutrino beam. Moreover, when reconstructing $\dptt$ there is a clear hydrogen peak but, even with an exceptional reconstruction, this is on top of a significant nuclear-target background which itself is quite sensitive to nuclear effects. $\dpt$ on the other hand predicts a vanishing nuclear-target background as $\dpt \rightarrow 0$.

A critical background when identifying neutrons coming from a neutrino interaction in some detector's active fiducial volume is from the neutrons coming from outside of its fiducial volume that enter in to it. These neutrons can be misidentified as primary neutrons if they are in coincidence with the interaction vertex time. However, neutrino beams operate with a time between bunches which is much longer than time between a neutrino interaction and a subsequent neutron scatter~\cite{Abe:2012av,dune-beam-1,dune-beam-2}. This means that the background neutrons will be delayed with respect to the beam bunch. Moreover, because external neutrons are produced without correlation to the primary neutrino interaction, the secondary neutrons arrive with a time and distance from the neutrino interaction vertex which is typically not similar from those of the signal neutrons. As a consequence, the contribution from these neutrons can be minimised by only considering those that fall within some range and time with respect to the primary neutrino interaction vertex. Moreover, background neutrons are not expected to be concentrated under the $\delta p_T$ hydrogen peak. A similar behavior is expected for the secondary neutron background, i.e. those produced by hadronic interactions of pions and protons within the active detector material.

\section{Conclusion}
\label{conclusion}

The ability to infer the true incoming neutrino energy from the products of a neutrino interaction is of crucial importance for current and future long-baseline neutrino oscillation experiments. However, current methods of mapping reconstructed to true neutrino energy tend to leave a wide smearing which has substantial dependence on the poorly understood details of neutrino-nucleus interactions.

In this manuscript a new method of anti-neutrino energy reconstruction is demonstrated, where a sub-sample of events quasi-free of the poorly understood nuclear effects, responsible for the wide neutrino energy smearing, is selected. This method relies on the identification of a sample of meson-less neutrino interactions producing at least one neutron in which there is very little imbalance between the outgoing lepton and neutron on the plane transverse to the incoming neutrino. This itself relies on the accurate reconstruction of neutron kinematics, which has been shown to be possible using a finely segmented plastic scintillator detector with three readout planes and fast timing. With realistic assumptions of potential detector performance it has been shown that such a sample of events can indeed be found and that these events are only marginally affected by the nuclear effects which are responsible for some of the dominant systematic uncertainties in current neutrino oscillation analyses. Overall, this method allows an interesting reduction of the strong correlations between the flux and the interaction models that arise in the measure of the neutrino oscillation probability. Similar results are expected for any detector with similar characteristics.

\section{Acknowledgements}
\label{sec:acknowledgements}

This work was initiated in the framework of the T2K Near Detector upgrade project, fruitful discussions in this context with our colleagues are gratefully acknowledged. S. Suvorov acknowledges support from RFBR grant \# 18-32-00070 and RSF Grant \# 19-12-00325. S. Dolan is acknowledges the support of a P2IO-CNRS grant. C. Jesús-Valls acknowledges support from MINECO and ERDF funds, Spain. 
 \\ % Cesar

\appendix

\section{Distance-energy correlations}
\label{app:leverarm}

\begin{figure}
\centering
\includegraphics[width=9cm]{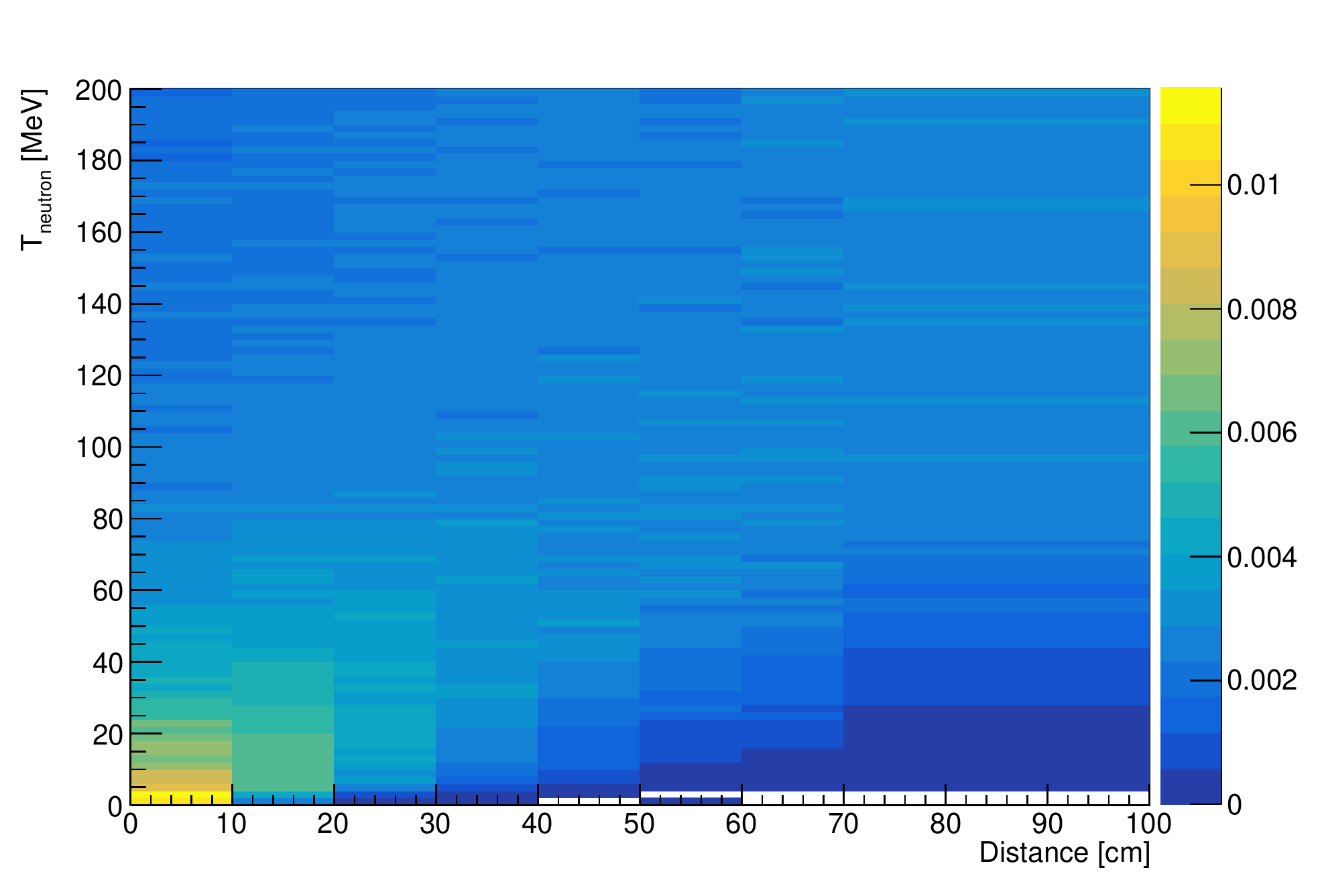}
\caption{\label{fig:neutron-leverarm} The relationship between neutron kinetic energy and travel distance through the scinitillator detector, determined from single particle GEANT4 simulations. Each slice in distance is normalised to unity.
}
\end{figure}

Neutrons typically travel a long distance before interacting. The resolution on the neutron ToF is driven by the detector time resolution, the neutron travel distance and the neutron speed. For instance, the measurement of the neutron ToF would be more precise if, given a certain travel distance, the neutron is slower or, given a certain speed, it travels further before interacting.  As shown in Fig.~\ref{fig:neutron-leverarm}, the neutron travel distance depends to some extent on its initial kinetic energy, in particular below 40~cm. This correlation is responsible for the feature shown in Fig.~\ref{fig:purEff}, where a stronger cut on the neutron lever arm reduces the anti-neutrino hydrogen interaction selection efficiency but it does not improve the selection purity, which is related to the neutron ToF resolution. In fact, the neutrons selected with a long lever arm are usually faster than neutrons selected with a shorter lever arm, resulting in a weak change of the neutron ToF resolution. Additionally, as discussed in Sec.~\ref{sec:results}, in selecting a region of lower neutron energies through requiring a particular lever arm, neutrino interactions on hydrogen can be disproportionately rejected (as these tend to be slightly higher energy). 

\section{Further analysis of neutrino energy resolution}
\label{app:selectedevents}

\begin{figure}
\centering
\includegraphics[width=10cm]{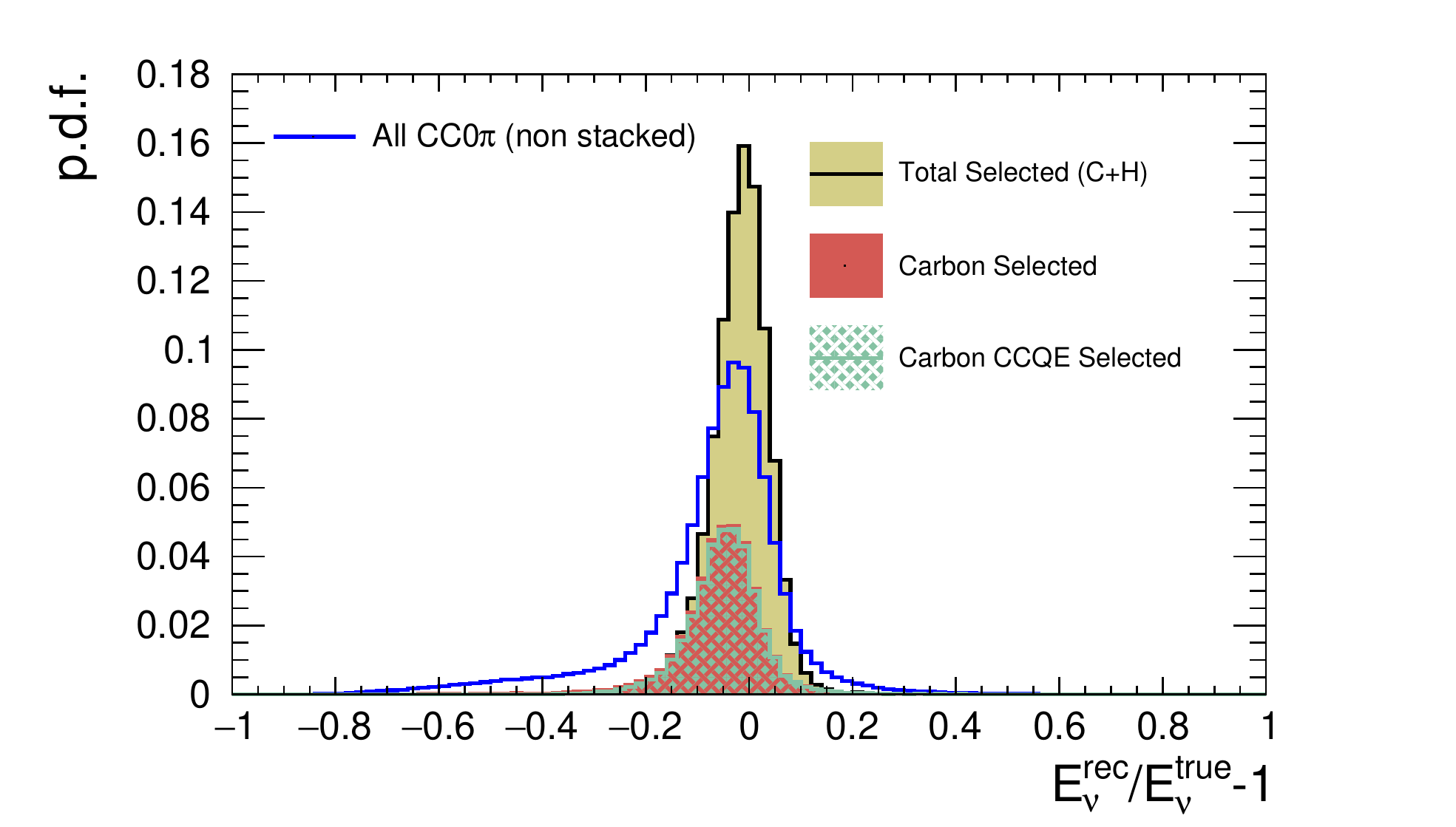}
\caption{\label{fig:recoStack} A one dimensional projection of Fig~\ref{fig:neutrino-energy-2D}, showing the neutrino energy resolution and bias before and after applying the $\dpt$ and lever arm cuts. The neutrino energy is reconstructed according to Eq.~\ref{eq:enu}. The events passing the cuts are split by interaction target and (for carbon) whether or not the selected events are from CCQE interactions. This figure is made considering the detector smearing and acceptances described in the text, using the timing resolution given by Eq.~\ref{eq:time-res-opt}.
}
\end{figure}

Fig~\ref{fig:recoStack} shows the neutrino energy resolution and bias before and after applying the $\dpt$ and lever arm cuts as described in~\ref{sec:analysis-strategy}, split by target nucleus and interaction mode. As also demonstrated in Fig.~\ref{fig:trueStack}, it can be seen that the selected carbon contribution is small and contributes a relatively small bias compared to carbon interactions found at larger $\dpt$. It is also clear that almost all of the selected carbon events are from CCQE interactions: less than 3\% of these are from other interaction modes, compared to 23\% without any $\dpt$ cut. 

\newpage

\bibliography{biblio}

\end{document}